# Best Practices for Fitting Machine Learning Interatomic Potentials for Molten Salts: A Case Study Using NaCl-MgCl$_2$

*Siamak Attarian [a*], Chen Shen [a*], Dane Morgan [a*], Izabela Szlufarska [a*]*

[a] Department of Materials Science and Engineering, University of Wisconsin, 1509 University Ave, Madison, WI, 53706, USA

* sattarian@wisc.edu

* cshen89@wisc.edu

* ddmorgan@wisc.edu

* szlufarska@wisc.edu

## Abstract

In this work, we developed a compositionally transferable machine learning interatomic potential using atomic cluster expansion potential and PBE-D3 method for $(NaCl)_{1-x}(MgCl_2)_x$ molten salt and we showed that it is possible to fit a robust potential for this pseudo-binary system by only including data from x={0, 1/3, 2/3, 1}. We also assessed the performance of several DFT methods including PBE-D3, PBE-D4, R2SCAN-D4, and R2SCAN-rVV10 on unary NaCl and MgCl$_2$ salts. Our results show that the R2SCAN-D4 method calculates the thermophysical properties of NaCl and MgCl$_2$ with an overall modestly better accuracy compared to the other three methods.

## 1. Introduction

Molten salts are materials with prospective applications in renewable energy systems such as heat transfer medium in molten salt reactors and heat storage materials in concentrated solar power plants [1–3]. Researchers are actively investigating the thermophysical and electrochemical properties of molten salts to find or design multicomponent salts that have a certain set of properties (ex. low melting point, low corrosivity, high specific heat capacity, low viscosity, etc.) tailored to certain applications [4–8]. As the main application of these salts is at high temperatures and impurities are difficult to control, the experimental characterization of the desired properties is challenging. Therefore, few experimental data are available and those mostly for unary salts (we will refer to salts by their number of cation components, e.g., NaCl is unary, NaCl-$MgCl_2$ is binary, etc.). Considering the vast design space of multicomponent salts, achieving a large database of the thermophysical properties of salts seems impractical through experimental means. Computational methods are inexpensive and faster alternatives to explore the properties of multicomponent salts. Among the available computational methods, *ab initio* molecular dynamics (AIMD) simulation using density functional theory (DFT) is the most accurate method commonly used to investigate some properties of molten salts [9,10]. Since AIMD calculation is computationally expensive, the sizes of the systems are usually limited to a few hundred atoms and the simulation times to a few tens of picoseconds, which allows one to calculate many properties, such as density, specific heat, etc., with acceptable uncertainties. However, for some properties, such as viscosity, the AIMD cannot reach adequate time scales to reduce uncertainties to an acceptable level for most applications. MD simulations with classical interatomic potentials [11,12] are low-cost options but the accuracies are questionable and fitting the parameters of the classical potentials is not an easy task.

Recent advances in machine learning interatomic potentials (MLIPs) have made it possible to run MD simulations of large systems for long time scales with *ab-initio* level accuracies [13,14]. MLIPs are also easy to fit as they mostly rely on robust optimization methods that have been well-developed in the computer science community. An important step in fitting MLIPs is the collection of fitting data. MLIPs are fitted to data calculated by *ab-initio* methods (usually DFT) and generating enough relevant DFT data that would fit a robust potential has been a concern since MLIPs were introduced. Specifically in the earlier attempts to fit MLIPs to molten salts, researchers have used formalisms based on feed forward neural networks which required tens of thousands of data points (where each data point is a DFT calculation of energy and forces (and sometimes stresses) for a specific configuration of atoms) [15,16]. It has been shown that linear MLIPs can fit potentials using less than a thousand training data [17]. Another aspect of training data generation is the relevancy of the data. Since DFT data generation is computationally expensive, researchers limit their data generation to the specific phenomena and the material that they are interested in studying. Almost all MLIPs that are developed for molten salts are either for a unary salt [15,16,18,19] or a few specific compositions of a binary or a ternary salt (mostly the eutectic point) [20–23][24], which makes the potential not transferable to the whole composition range between the end member salts. A few cases that fitted potentials across a range of compositions [25,26] have used up to 10 distinct compositions between the end members to achieve transferability, which can make the data generation a time-consuming and complicated task, especially for multicomponent salts where the composition space expands significantly with the number of components. One aim of this study is to determine the minimum number of compositions to include in the training data to achieve a transferable potential for a binary molten salt, at least for one representative example.

Another interesting point in MLIP development is their accuracy. The accuracy of an MLIP with respect to DFT is assessed by various means, most notably by evaluating the energy/force errors using an independent testing set. However, in the end, the predictions of MD simulations with MLIPs should be validated by experimental results, and for a well-fitted potential, the deviation of the predictions from the experiments may be attributed either to the uncertainties of the experimental results, statistical uncertainties of MD simulations, or inaccuracy of the underlying DFT method. Some of the first AIMD studies in molten salts demonstrated that including dispersion effects in DFT increases the accuracy of the calculated properties [27] and including dispersion effects have become a routine practice in modeling of these materials [10,28,29]. With recent advances in exchange-correlation formulations such as SCAN [30], R2SCAN [31], etc., and newly introduced dispersion correction methods such as DFT-D4 [32], it is interesting to study the accuracy of these new variants in modeling molten salts. As many of these new variants are tested on properties such as lattice constants, formation energies, elastic constants, etc., for crystalline phases at 0 K, their performance in predicting thermophysical properties, especially in molten phases, is unknown. Therefore, a second aim of this study is to compare the accuracy of some of these DFT+dispersion methods in predicting the thermophysical properties of molten salts by using MLIPs as a surrogate to DFT.

In this work, we chose $NaCl-MgCl_2$ binary salt and developed a transferable potential for this system. There are two MLIPs available in the literature for this salt both of which only work for the eutectic composition ($(NaCl)_{0.58}(MgCl_2)_{0.42}$) [20,33]. We used the Atomic Cluster Expansion method (ACE potential) [34] for the fitting as it showed good performance in our previous work [24] and used PBE-D3 [35] as our primary DFT method. During the development of the potential, we used different training sets composed of training data from various compositions and determined the lowest number of compositions needed for a compositionally transferable potential, and compared the property predictions between them. We also ran several AIMD simulations of $NaCl$, $MgCl_2$, and eutectic $NaCl-MgCl_2$ to validate the ACE predicted properties. These showed such good agreement that we concluded that the ACE potential can be used as a surrogate to effectively assess what would be obtained from AIMD for different exchange-correlation formulations. We then took advantage of this capability and compared different exchange-correlation formulations by recalculating the same training data using PBE-D4 [32], R2SCAN-D4 [36], and R2SCAN-rVV10 [37,38] methods, fitting multiple ACE potentials, and comparing the predicted material properties to the experimentally available values.

## 2. Methodology

### 2.1. Fitting the ACE potentials

We start by generating data and fitting potentials for the end member salts, NaCl and $MgCl_2$, and then for the atomic configurations in between them $((NaCl)_{1-x}(MgCl_2)_x)$. For the rest of this article, the terms atomic configuration and training data are used synonymously. Each training data consists of a cell of atoms (usually around 100 atoms) and its DFT calculated energy and atomic forces. Here we explain the process of data generation for NaCl. The same procedure is carried out for $MgCl_2$. There are many approaches to generate training data for a system. The methods based on active learning are popular as they efficiently generate required data for training a robust potential [25,39–42]. Active learning or active sampling is a procedure in which the raw atomic configurations (configurations that are not DFT calculated yet) are generated and a subset of them are collected based on some specific criterion. The pyace package [43] that we used to fit ACE potentials in this work and its LAMMPS [44] library includes an active learning method based on the D-optimality criterion [39]. The details of the ACE potential formalism and D-optimality criterion have been extensively discussed in the main reference papers and our previous works [17,24,39,40,42]. In the following we will, for convenience, use the phrase "DFT calculate" to refer to calculating a structure with DFT. Below we describe the general procedure of active learning:

1- Generate a set of initial data and DFT calculate them and construct the first training set.

2- Fit an ACE potential using the training set.

3- Run MD simulations using the fitted ACE potential and generate more raw atomic configurations. Check the D-optimality criterion on the raw atomic configurations and select the next set of training data.

4- DFT calculate the selected training data. Add them to the training set.

5- Repeat steps 2-4 until the D-optimality does not select any raw atomic configurations or the number of the selected configurations is much lower than the total number of configurations in the MD trajectory (less than 1% for example). Each iteration of steps 2-4 is called one active learning cycle.

To generate the initial training data in step 1, we ran MD simulations with Fumi-Tosi potential [45] using LAMMPS software. The simulation started from the crystalline phase of NaCl using a simulation cell with 100 atoms and was performed in the temperature range of 0 K to 1600 K and pressure range of -1 GPa to 1 GPa overall for 1 ns and eventually 100 atomic configurations were collected from the trajectory of this simulation at every 10 ps. These atomic configurations were DFT calculated and were then used to train an ACE potential in step 2. The main hyperparameters of the ACE potential include the cutoff radius, the number of many-body interactions, the number of radial basis functions and the number of angular basis functions, and they can be separately tuned for each many-body interaction between atomic species. We used several different sets of hyperparameters and evaluated their energy/force errors and their computational cost and in the end, we chose the hyperparameters provided in Table 1. One noteworthy finding of this hyperparameter tuning was that for a good description of ionic bonding between Na, Cl (and later

Mg) one needs to use up to 4 body interactions. For 5- and higher body interactions there is not any noticeable increase in the accuracy of the potential.

In step 3 we ran MD simulations with the developed potential in the temperature range of 0 K to 1600 K and pressure range of -1 GPa to 1 GPa for 1 ns. Since the active sampling function of ACE potential is implemented in LAMMPS, the D-optimality check is done during the MD simulation at each time step with a parameter called extrapolation grade. If the extrapolation grade of an atomic configuration is larger than 1, it means that the current training set cannot confidently predict the energy/forces of this atomic configuration, so it is a good candidate to be added to the next training set. Usually, in the first few cycles of active learning, thousands of configurations may be selected as good candidates but due to the similarity (correlation) of many of these configurations, we do not pick and DFT calculate all of them. At each cycle, we chose 200 configurations and DFT calculated them in step 4 and repeated the cycle. It is worth mentioning that at each active learning cycle, we perform the MD simulations with a newly developed ACE potential from the previous step, and although the simulation starts with the same initial configuration, eventually its trajectory will be different than the previous MD simulations and the system will visit different microstates, which adds diversity to our training data. After a few active learning cycles, we ended up with 841 training data for NaCl. The same procedure was carried out for $MgCl_2$ and in the end, 814 training data were collected. We name the final ACE potentials that were fitted for these two unary data sets Pot_N and Pot_M for NaCl and $MgCl_2$ respectively.

For the binary system, there could be infinite possible compositions of NaCl-$MgCl_2$. To limit the composition space, we kept the number of atoms between 98 and 100 in each cell and this gave us 32 compositions ranging from $(NaCl)_{0.98}(MgCl_2)_{0.02}$ (48 Na, 1 Mg, 50 Cl) to $(NaCl)_{0.03}(MgCl_2)_{0.97}$ (1 Na, 32 Mg, 65 Cl). Then we fitted 4 potentials. The first potential (Pot_1) included training data from the end members (841 NaCl and 814 MgCl2) and one composition in the middle ($(NaCl)_{1-x}(MgCl_2)_x$ where x = {0, 0.5, 1}). The second potential (Pot_2) included data from the end members and two compositions ($(NaCl)_{1-x}(MgCl_2)_x$ where x = {0, 0.33, 0.67, 1}). The third potential (Pot_3) included data from the end members and three compositions ($(NaCl)_{1-x}(MgCl_2)_x$ where x = {0, 0.25, 0.5, 0.750, 1}). The fourth potential (Pot_4) included training data from the end members and all the 32 compositions in between. To train each potential we used a procedure similar to what was performed for NaCl. For example, for the training of Pot_3 we started by running 3 separate MD simulations at each of the 3 mentioned compositions and picked 33 configurations from each composition, DFT calculated them and, put them together along with the end member data, fitted an ACE potential and then repeated a few active learning cycles until there were no new training data needed to be added. The initial configurations for each composition were created using PACKMOL code [46].

We also generated an independent testing data set by running MD simulations of each of the 34 compositions (2 end members and 32 in between) at the same range of pressures and temperatures that the training data were produced and collected at least 50 configurations at each composition, with the configurations widely separate in time compared to the correlation time of the atomic configurations. The primary DFT method used for generating the data was PBE-D3. In the end, we recalculated the DFT data of NaCl and MgCl2 using PBE-D4, R2SCAN-D4, and R2SCAN-rVV10 methods, refit separate ACE potentials, ran one more active learning cycle to make sure of the robustness of the potentials, and used them to compare different DFT approaches.

All the DFT calculations were performed using VASP 6.4.2 package [47]. The PAW-PBE potentials that were used in this study are Na_pv ($2p^63s^1$), Mg ($3s^2$), and Cl ($3s^23p^5$). An energy cutoff of 600 eV was used for the plane-wave basis set and a single gamma point was used to sample the Brillouin zone.

## 2.2. MD simulations

To compare the performance of the potentials with respect to DFT, we ran AIMD simulations of NaCl, $MgCl_2$, and the eutectic salt ($(NaCl)_{0.58}(MgCl_2)_{0.42}$) at temperatures 1200 K and 1500 K and at pressure $P = 0$ GPa for 50 ps with a time step of 1 fs using PBE-D3 method with simulation cells with 100, 99 and 116 atoms respectively using VASP. Using these simulations, we calculated radial distribution function (RDF), density, average specific heat, and average thermal expansion coefficient for the temperature range 1200 K - 1500 K.

All MD simulations (ACE MD) in this work are performed using LAMMPS code. All simulations had a time step of 1 fs. To calculate density, thermal expansion coefficient, specific heat, and diffusivity we ran MD simulations at controlled pressure – controlled temperature (NPT) ensemble at $P = 0$ GPa and the temperature range 1100 K $< T <$ 1500 K for 100 ps with simulation cells containing around 6000 atoms. At each pressure and temperature, we ran three independent MD simulations, calculated the desired property, and reported the average value and the error in the mean. We performed these sets of simulations for NaCl, $MgCl_2$, and binary salts at three compositions $(NaCl)_{1-x}(MgCl_2)_x$ where x = {0.62, 0.42, 0.24} (compo1, compo2, and compo3 respectively) since there are experimental values available for the density of these compositions to compare to the MD results. The density $\rho$ is calculated as the ensemble average of the density at each temperature. The thermal expansion coefficient at each temperature is calculated by:

$$\alpha = -\frac{1}{\rho}\left(\frac{d\rho}{dT}\right)\Big|_p \qquad Eq.\,1$$

and the specific heat capacity at each temperature is calculated by:

$$c_p = \frac{\partial h}{\partial T}\Big|_p \qquad Eq.\,2$$

where $h$ is the ensemble average of the enthalpy at each temperature. The self-diffusion coefficient is calculated from the slope of mean squared displacement (MSD) obtained from the LAMMPS code, using Einstein's relation [48]:

$$D = \frac{1}{6}\lim_{t\to\infty}\frac{d}{dt}\left[\frac{1}{N}\sum_{i=1}^{N}(r_i(t) - r_i(0))^2\right] \qquad Eq.\,3$$

where $D$ is the self-diffusion coefficient, $N$ is the number of atoms, and $r_i(t)$ is the position of atom $i$ at time $t$.

For viscosity calculations, we used the Green-Kubo relation [49,50]:

$$\eta = \frac{V}{k_B T}\int_0^{\infty} <P_{\alpha\beta}(t).P_{\alpha\beta}(0)> dt \qquad Eq.\,4$$

Where $\eta$ is the viscosity, $k_B$ is the Boltzmann constant, and $P_{\alpha\beta}$ are the off-diagonal elements of the stress tensor. In each viscosity simulation after an initial 30 ps equilibration of the system in the NPT ensemble, we ran the simulation for another 5 ns in NVE ensemble to calculate the viscosity. An autocorrelation time of 20 ps was chosen for these simulations which was enough to decay the autocorrelation function of the diagonal stress components to zero. For each viscosity calculation, we ran 3 separate simulations starting from different atomic configurations, and in the end, we reported the average value and standard deviation of the 3 runs as the error. We performed viscosity calculations for NaCl, MgCl$_2$, and binary salts at three compositions (NaCl)$_{1-x}$(MgCl$_2$)$_x$ where x = {0.70, 0.50, 0.30} (compo4, compo5, and compo6 respectively) since there are experimental values available for the viscosity of these compositions to compare to the MD results. These simulations were performed at various temperatures between 973 K to 1300 K for different systems. Each system had around 3000 atoms.

## 3. Results

### 3.1 ACE potentials

We fitted several ACE potentials using the data generated by PBE-D3 approach as discussed in section 2.1. The hyperparameters of the potentials for each pair of species are provided in Table 1.

*Table 1. Hyperparameters of each pair of species in the ACE potentials fitted in this work for various systems.*

|  | Na-Na | Mg-Mg | Cl-Cl | Na-Cl | Mg-Cl | Na-Mg |
|---|---|---|---|---|---|---|
| Pot_N | n=15, 1, 1<br>l= 0, 1, 1 | - | n=15, 1, 1<br>l= 0, 1, 1 | n=15, 1, 1<br>l= 0, 1, 1 | - | - |
| Pot_M | - | n=15, 3, 1<br>l= 0, 1, 1 | n=15, 3, 1<br>l= 0, 1, 1 | - | n=15, 3, 1<br>l= 0, 1, 1 | - |
| Pot_1 to Pot_4 | n=15, 1, 1<br>l= 0, 1, 1 | n=15, 3, 1<br>l= 0, 1, 1 | n=15, 3, 2<br>l= 0, 2, 2 | n=15, 1, 1<br>l= 0, 1, 1 | n=15, 3, 1<br>l= 0, 1, 1 | n=15, 3, 1<br>l= 0, 1, 1 |

In the ACE potential, each pair (and triplet, etc.) of atomic species, has its own sets of hyperparameters which in turn are divided into two-body, three-body, four-body, etc., parameters.

In Pot_1 to Pot_4 for the triple interaction (Na-Mg-Cl which is not shown in the table) we used hyperparameters the same as Cl-Cl.

Table 2 shows the energy/force errors of the training set for each potential.

*Table 2. Energy and force errors of the training data of each potential. The number of training data for potentials Pot_1 to Pot_4 also includes the training data of unary NaCl and MgCl$_2$.*

|  | Number of training data | Energy (meV/atom) | Force (Na) (meV/Å) | Force (Mg) (meV/Å) | Force (Cl) (meV/Å) |
|---|---|---|---|---|---|
| Pot_N | 841 | 1.1 | 23 | - | 24 |
| Pot_M | 814 | 2.4 | - | 68 | 59 |
| Pot_1 | 2187 | 2.1 | 24 | 67 | 48 |
| Pot_2 | 2283 | 2.0 | 25 | 67 | 48 |
| Pot_3 | 2278 | 2.0 | 25 | 67 | 48 |
| Pot_4 | 2613 | 2.1 | 25 | 68 | 50 |

Overall, our results in Tables 1 and 2 suggest that Na ion is an easier species to fit compared to Mg ion since fewer parameters are needed to fit the Na ion and they also achieve lower errors. Using a testing set containing 1982 atomic configuration from various compositions of NaCl-MgCl$_2$ we tested the performance of Pot_1 to Pot_4 as shown in Figure 1. In Figure 1(a) we compare the energy errors of Pot_1 to Pot_4 and in Figure 2(b) we remove the results of Pot_1 for a better comparison between the energy errors of Pot_2 to Pot_4. The leftmost composition (composition #1) is pure NaCl, and the rightmost composition (composition #34) is pure MgCl$_2$. The list of compositions is provided in the supplementary materials (Table S.1). As can be seen in Figure 2(a), Pot_1 does not perform well in predicting energies across the compositions and only works well for the two end members and in the middle where similar data are provided in the training set. In Figure 2(b) we see that the energy errors of Pot_2, Pot_3, and Pot_4 are very close for the whole composition range. In Figure 2(c) we compare the total force errors of Pot_1 to Pot_4. The figure shows that the force errors are close across the compositions, especially for the right half of the plot. It is worth mentioning that there is a difference between the way energies and forces are fitted in MLIPs. Since the atomic forces are directly accessible from DFT, the MLIP is directly trained on the forces of each atom, but for the energies we only have access to the total energy of each atomic configuration, and the potential learns the energy contribution of each atom such that the sum of the energies would be equal to the total energy of the system. This could be why ACE learns the forces better than the energies for Pot_1 despite the less diversity in the atomic environments. Another aspect of Figure 1(c) is the increase in the force errors from left to right. As shown in Table 2, the force errors of Mg ions are higher than Na ions, and in Figure 1(c) the more we go to the right of the plot the more Mg ions are present in the configuration. Overall, Figure 1 suggests that including data from only two compositions between the end members provides sufficient diversity in the atomic environments to achieve a compositionally transferable potential for binary molten salts.

Figure 2 compares the predicted densities of NaCl (Fig 2(a)) and $MgCl_2$ (Fig 2(b)) in the temperature range 1100 K to 1500 K by Pot_N, Pot_M, and Pot_2 to Pot_4 and the calculated densities from AIMD. The first thing to notice is that the predicted densities from different potentials do not overlap, and they have up to 1% difference. In the case of NaCl, the predictions of Pot_2 to Pot_4 which are trained on the data of unary and binary salts overlap, while the predictions of Pot_N which is only trained on NaCl data slightly differ. In $MgCl_2$, all the predictions from the potentials slightly differ from each other. The uncertainty from the ensemble averaging of each MD simulation is on the order of 0.001 g/cm³, which is too small to be detectable in the graph, so we did not plot them, and it is unlikely to be the reason for this discrepancy.

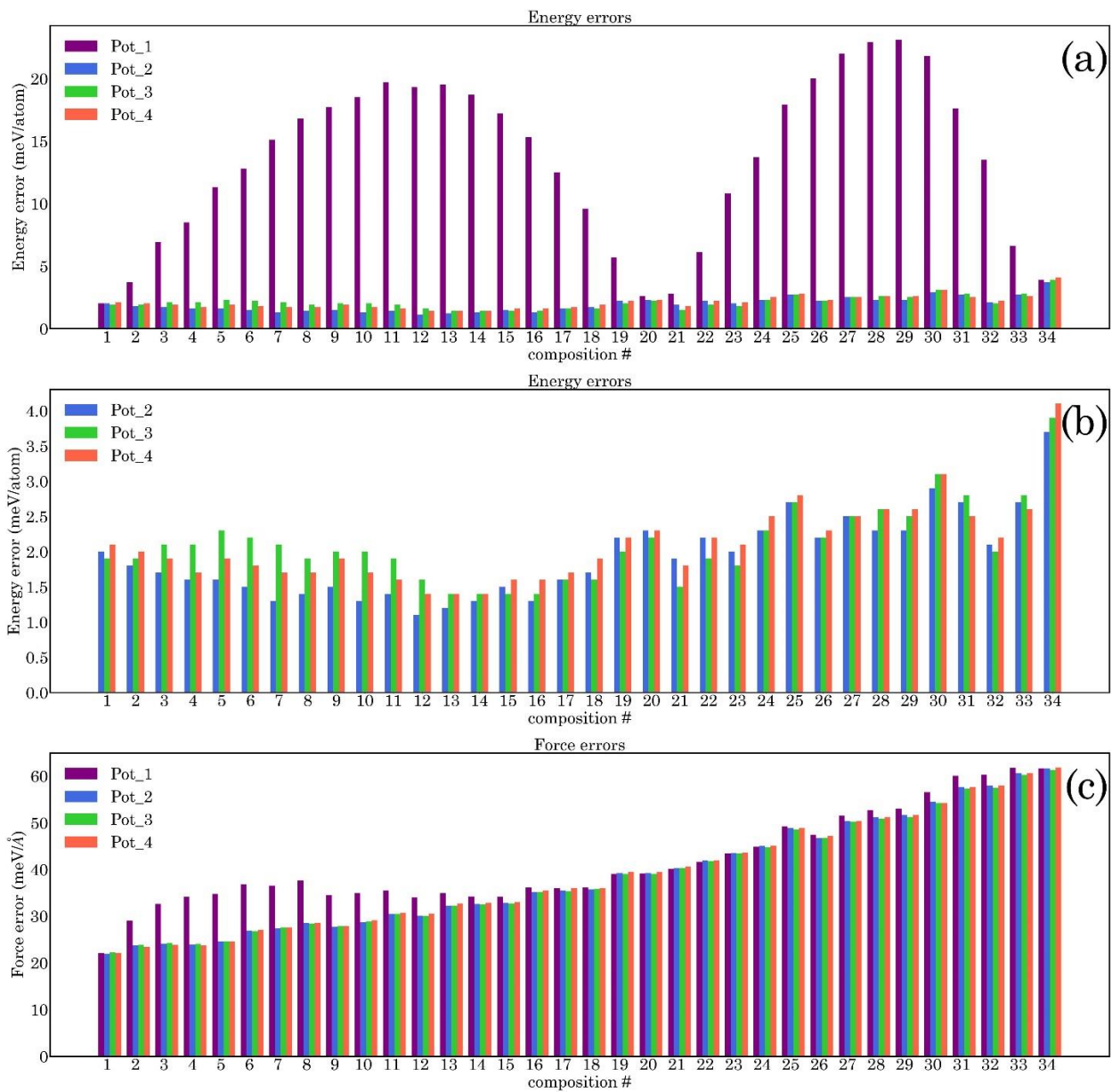

Figure 1. Energy errors of (a) Pot_1 to Pot_4, (b) pot_2_pot_4, and (c) force errors of Pot_1 to Pot_4 across the compositional space between NaCl (#1) and MgCl$_2$ (#34). Plot (b) is the same as plot (a) by excluding the errors of Pot_1 for a better comparison of the Pot_2 to Pot_4

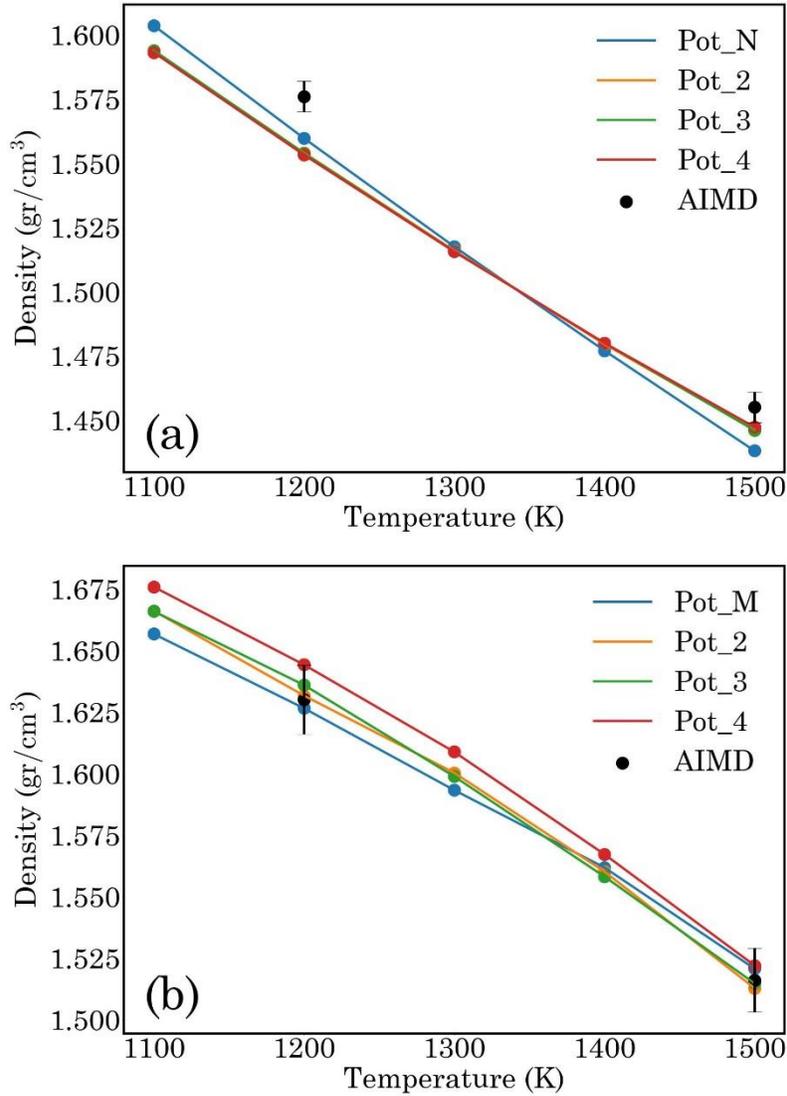

*Figure 2. Densities of (a) NaCl, and (b) MgCl$_2$, calculated using MD simulations with Pot_M, Pot_N, and Pot_1 to Pot_4. The AIMD calculated data are also shown in the plots with their error bars*

We believe the main reason for such discrepancies is the uncertainty that comes from fitting a potential to DFT data. While generally the uncertainties of potentials are reported in terms of energy/force errors, these values do not directly determine the uncertainties in the predicted properties. In the supplementary materials we show that a 2 meV/atom error in the energy of NaCl could result in about 1.1% uncertainty in the calculation of the density compared to the reference method (DFT in this case). Such uncertainties in property prediction are part of the fitted potentials when the energy/force errors are not exactly zero and for two potentials with the same (or close) root mean squared errors (RMSE) of energies and forces, one cannot guarantee that both will calculate the same value for a certain property at a certain pressure and temperature. Overall, by the results of Figures 2 and 3, we conclude that Pot_2 is as good as other potentials, and we will use Pot_2 for the rest of this work and we will call it simple "ACE" when comparing to AIMD results. The parity plots of energies and forces of the testing set, predicted by the aforementioned potentials, are provided in the supplementary materials (Figure S1). Figure 3 compares the predicted densities of ACE with AIMD calculated densities. For NaCl, the density prediction of ACE is around 1% lower than AIMD at 1200 K and is around 0.5% lower at 1500 K. For the Eutectic salt (which was not included in the training data) the predictions are about 0.5% lower than AIMD and within the uncertainties at both temperatures and for $MgCl_2$ the predicted density of ACE is almost the same as AIMD.

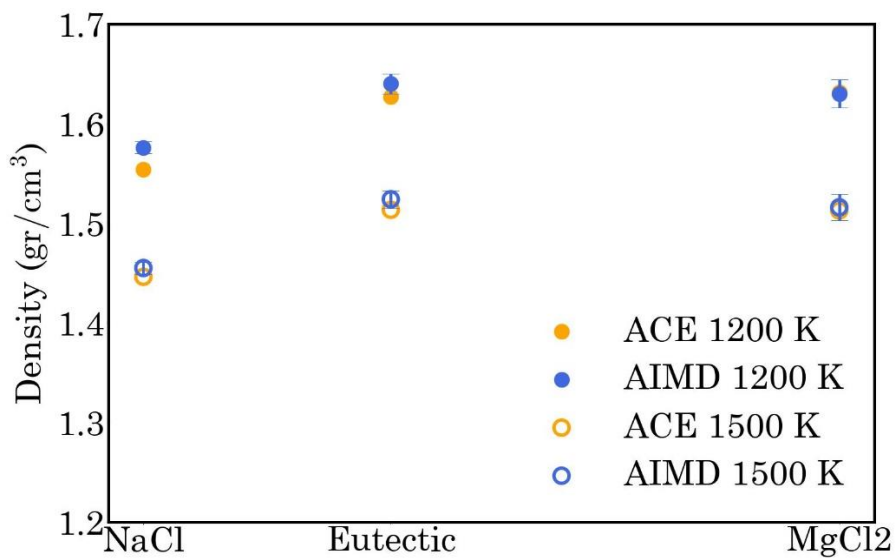

Figure 3. Comparison between ACE predicted densities and AIMD predicted densities of NaCl, MgCl$_2$, and the eutectic NaCl-MgCl$_2$ ((NaCl)$_{0.58}$(MgCl$_2$)$_{0.42}$) at 1200 K and 1500 K

Figure 4 compares the radial distribution function (RDF) of Na-Cl, Mg-Cl, and Na-Mg pairs in the eutectic salt calculated by ACE and AIMD at (a) 1200 K and (b) 1500 K. All the curves from both temperatures match nearly perfectly. The RDF comparison of NaCl and MgCl$_2$ systems is also provided in the supplementary materials (Figures S2 and S3).

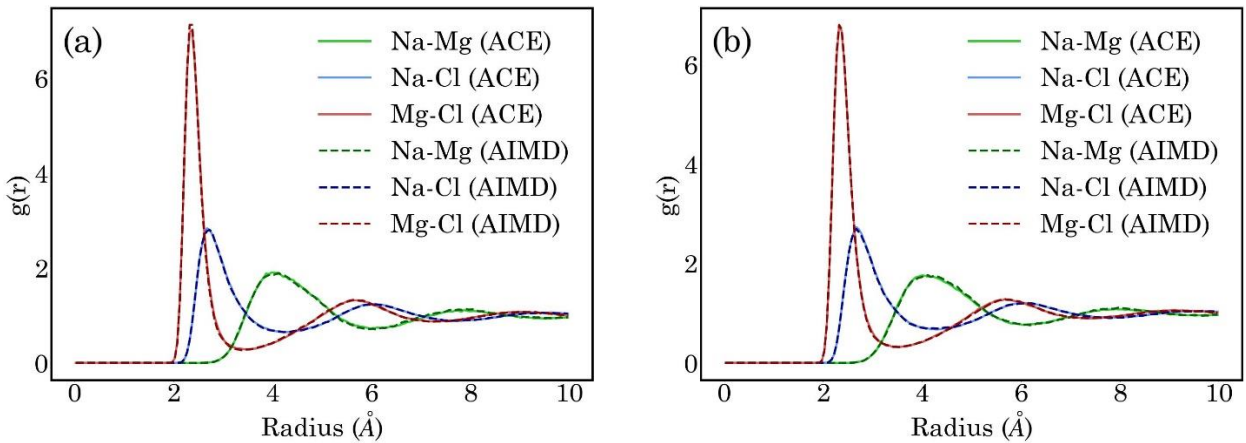

Figure 4. Radial distribution functions (RDF or g(r)) of Na-Cl, Mg-Cl, and Na-Mg pairs in the eutectic NaCl-MgCl$_2$ salt at temperatures (a) 1200 K, and (b) 1500 K calculated from ACE and AIMD

In Table 3 we compare the average specific heat and average thermal expansion coefficient of NaCl, MgCl$_2$, and eutectic salt calculated by ACE and AIMD between temperatures 1200 K and 1500 K. These values are calculated by Equations 1 and 2 based on the slope between two data points. The density in the denominator of Equation 1 is taken to be the average density between 1200 K and 1500 K. The uncertainties of the ACE calculations are negligible (at most 0.1 J/mol.K and on average 0.06 J/mol.K for the specific heat and at most 0.02×10$^{-4}$ 1/K and on average 0.01×10$^{-4}$ 1/K for the thermal expansion coefficient), so we did not show them in the table.

*Table 3. Comparison of the average specific heat and average thermal expansion coefficient (TEC) of NaCl, MgCl$_2$, and the eutectic salt calculated by ACE and AIMD between 1200 K and 1500 K. Uncertainties are one standard deviation in the given value. The uncertainties of the ACE calculations are negligible (at most 0.1 J/mol.K and on average 0.06 J/mol.K for the specific heat and at most 0.02×10$^{-4}$ 1/K and on average 0.01×10$^{-4}$ 1/K for the thermal expansion coefficient).*

|  | ACE | AIMD |
|---|---|---|
| Specific heat (J/mol.K) |  |  |
| NaCl | 62 | 63 ± 4 |
| MgCl2 | 97 | 87 ± 8 |
| Eutectic | 77 | 81 ± 5 |
| TEC (1/K) |  |  |
| NaCl | 2.4 × 10$^{-4}$ | 2.7 ± 0.2 × 10$^{-4}$ |
| MgCl2 | 2.5 × 10$^{-4}$ | 2.4 ± 0.4 × 10$^{-4}$ |
| Eutectic | 2.4 × 10$^{-4}$ | 2.4 ± 0.3 × 10$^{-4}$ |

The specific heat of NaCl calculated by ACE is very close to and within the uncertainties of AIMD. For MgCl$_2$, ACE predicts a specific heat 1% higher than the higher bound of AIMD predictions from one standard deviation but well within two standard deviations. For the eutectic, the ACE prediction is also within the uncertainties of AIMD. In the case of the thermal expansion coefficients for NaCl, the ACE prediction is 3% lower than the lower bound of AIMD predictions for one standard deviation but well within two standard deviations. For eutectic salt and MgCl$_2$, the predicted values are within the one standard deviation of the AIMD. Overall, the fitted ACE potential shows a very good performance in replicating AIMD results and we used the potential to calculate densities and viscosities across the compositions at various temperatures. For density calculations, we performed our simulations in the temperature range of 1000 K to 1500 K. Figure 5 shows the predicted densities of (a) NaCl, (b) MgCl$_2$, (c) compo1, (d) compo2 and (e) compo3 ((NaCl)$_{1-x}$(MgCl$_2$)$_x$ where x = {0.62, 0.42, 0.24} respectively) and compares them to the experimental results [51]. The experimental results are available for a narrow temperature range. The ACE potential (which is a surrogate for PBE-D3) overestimates the density of all systems. While it shows a good performance for $MgCl_2$ with less than 1% error, at the highest, it has up to 4% error in predicting the density of NaCl. Figure 6 shows the density of the NaCl-$MgCl_2$ system across the compositions. In Figure 6(a) we compared the predicted density with the experimental results at 1100 K where there are available results for all the systems. In Figure 6(b) we show the predictions of ACE across the compositions in the temperature range of 1000 K to 1500 K. In all cases, we predict a significant upward bowing of the density vs. a simple linear interpolation. When experimental data is unavailable, figures such as Figure 6(b) are a useful resource for estimating the density of a desired composition of the binary salt at the desired temperature.

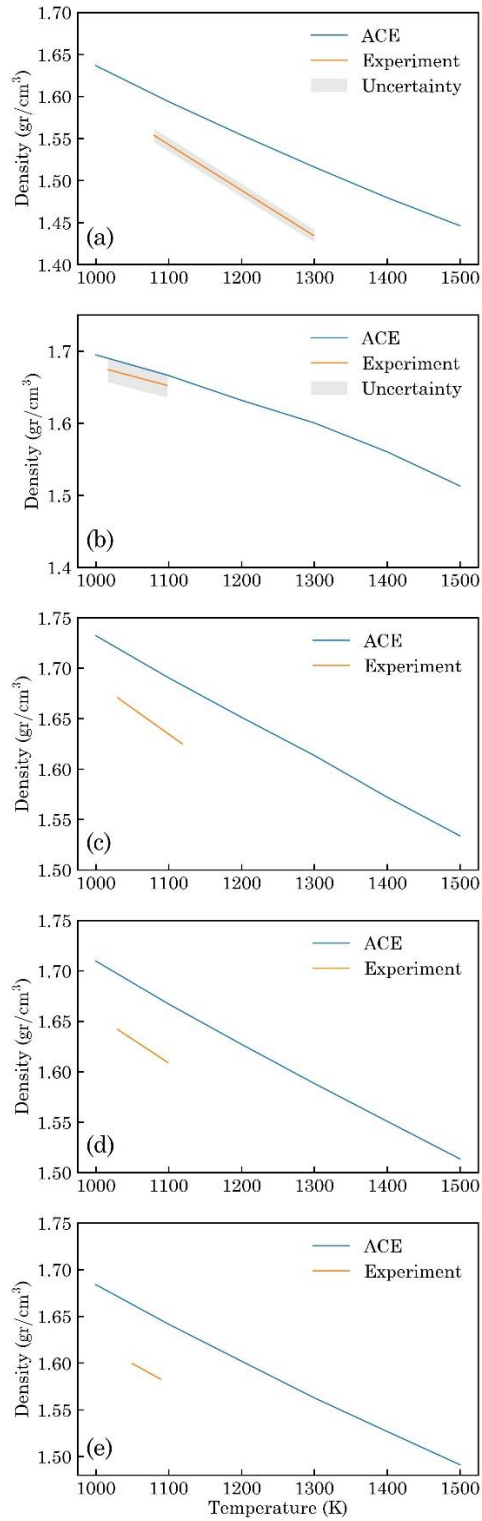

*Figure 5. Densities of (a) NaCl, (b) MgCl$_2$, (c) compo1 ((NaCl)$_{0.38}$(MgCl$_2$)$_{0.62}$), (d) compo2 ((NaCl)$_{0.58}$(MgCl$_2$)$_{0.42}$), and (e) compo3 ((NaCl)$_{0.76}$(MgCl$_2$)$_{0.24}$)*

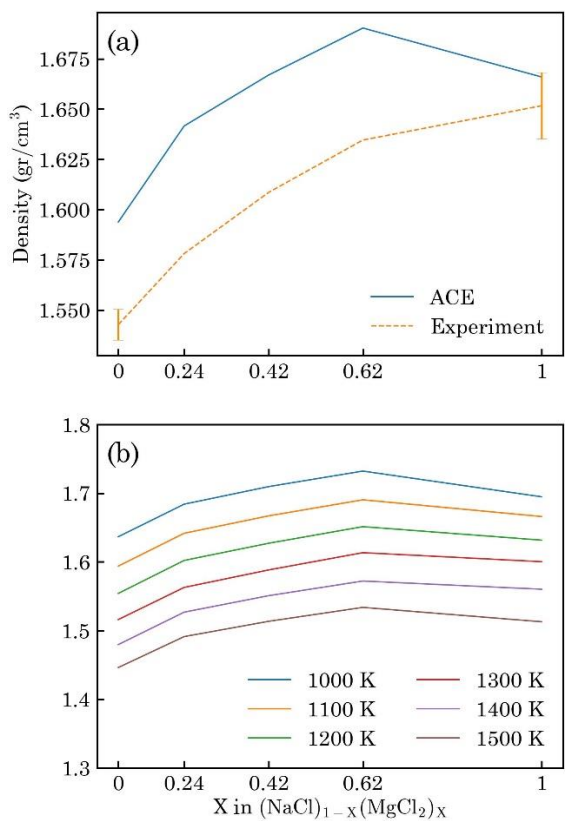

*Figure 6. Densities across the compositions of NaCl-MgCl$_2$: (a) ACE vs experiment at 1100 K and (b) ACE predictions in the temperature range 1000 K to 1500 K. The left most point is NaCl and the right most point is MgCl$_2$.*

Figure 7 compares the viscosities calculated by ACE for (a) NaCl, (b) MgCl$_2$, (c) compo4, (d) compo5, and (e) compo6 ((NaCl)$_{1-x}$(MgCl$_2$)$_x$ where x = {0.70, 0.50, 0.30} respectively) and experimental results [51]. The first thing to notice in this plot is the uncertainties in the calculation of the viscosities, which in some cases are up to 10%. We originally ran our viscosity calculations for 5 ns, and we observed that after the first 3 ns, the initial instability of the Green-Kubo integral fades away but during the rest of the simulation the viscosity has minor fluctuations. We let many of these simulations continue to run up to 10 ns, but the viscosity never fully converged and had minor oscillations. To get a better estimate, we ran three separate simulations for 5 ns at each temperature, each starting from a different initial state, and took the average of the three. The reported uncertainties are the standard deviation of the three simulations. The second thing to notice is the overall deviation of the calculated viscosities from experimental results. For NaCl, the viscosities are overestimated by up to 10% while for $MgCl_2$ the viscosities are underestimated by up to 20%. This is the opposite of the density predictions where the densities of $MgCl_2$ had been calculated more accurately compared to NaCl. For compo4, compo5, and compo6, the predictions are closer to the experimental values which could be due to the errors of NaCl and $MgCl_2$ correcting each other. It should be noted that we only found one set of experimental data and the stated uncertainties for these data are reported as 1%. The availability of different viscosity measurements would have allowed a better assessment of the accuracy of the potential and the accuracy of PBE-D3. In the supplementary materials, we have also provided a comparison between the viscosity prediction between Pot_2 and Pot_3, which illustrates that there is no statistically significant difference in the viscosity predictions of the two potentials (Figure S4).

In the next section, we present the results of our calculated thermophysical properties of salts using the more recent methods of PBE-D4, R2SCAN-D4, and R2SCAN-rVV10. Due to the time-consuming nature of many of the simulations done in this work, especially viscosity calculations, we limit our comparisons to unary NaCl and $MgCl_2$ salts.

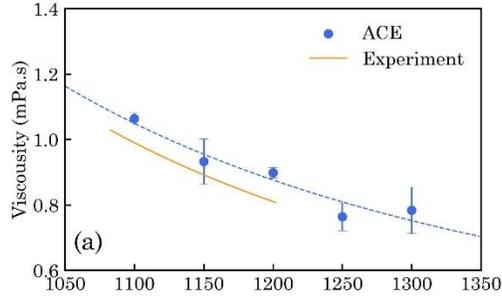

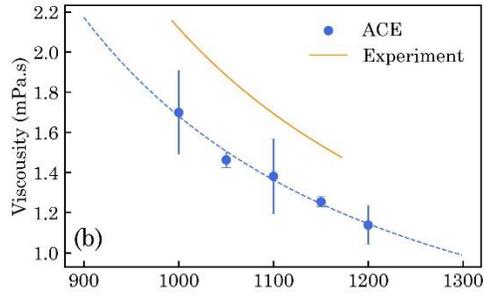

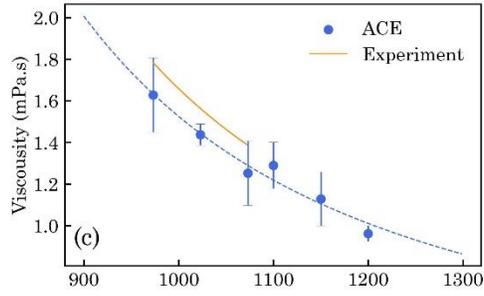

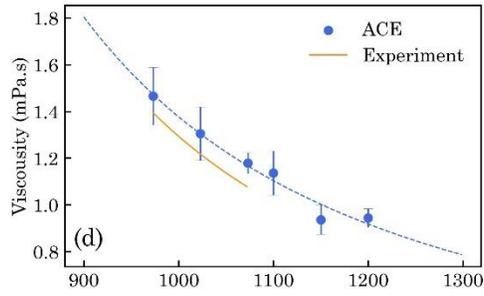

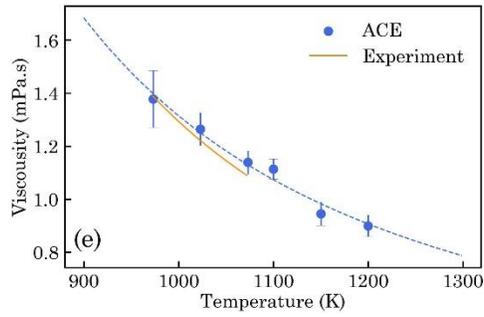

*Figure 7. Viscosities of (a) NaCl, (b) MgCl$_2$, (c) compo4 ((NaCl)$_{0.3}$(MgCl$_2$)$_{0.7}$), (d) compo5 ((NaCl)$_{0.5}$(MgCl$_2$)$_{0.5}$), and (e) compo6 ((NaCl)$_{0.7}$(MgCl$_2$)$_{0.3}$)*

## 3.2 DFT method comparison

To fit new potentials to NaCl and MgCl$_2$, we took the configurations that we generated during the fitting of PBE-D3, re-calculated them by other DFT methods, and fitted new potentials. We also performed additional active learning cycles for each new potential to include more configurations for a better fit and increased the hyperparameters as needed to reach good accuracy. Table 4 shows the accuracy of the newly fitted potentials.

*Table 4. Energy and force errors of the training data of each potential.*

| Potential \ Error | Number of training data | Energy (meV/atom) | Force (Na) (meV/Å) | Force (Mg) (meV/Å) | Force (Cl) (meV/Å) |
|---|---|---|---|---|---|
| NaCl | | | | | |
| PBE-D3 | 841 | 1.1 | 23 | - | 24 |
| PBE-D4 | 851 | 1.6 | 181 | - | 100 |
| R2SCAN-D4 | 853 | 1.5 | 69 | - | 45 |
| R2SCAN-rVV10 | 927 | 1.9 | 34 | - | 35 |
| MgCl$_2$ | | | | | |
| PBE-D3 | 814 | 2.4 | - | 68 | 59 |
| PBE-D4 | 828 | 2.1 | - | 62 | 52 |
| R2SCAN-D4 | 829 | 2.1 | - | 62 | 53 |
| R2SCAN-rVV10 | 828 | 2.0 | - | 60 | 52 |

For NaCl, especially the PBE-D4 method, we faced problems in fitting the potential. The generated data using the PBE-D4 method seemed to have some irregularities and the calculated forces on some atoms were not trainable, even with increasing hyperparameters their error would not decrease, and they acted more like outlier data. Regardless, more than 70% of the atoms have force errors < 40 meV/Å and 90% have force errors <100 meV/Å. We have more discussion of these fitting issues in the supplementary materials (Figures S5 to S8 and the discussions related to them), and we believe that the mentioned issue stems from the DFT calculation method (or code) rather than MLIP fitting. Figure 8 compares the densities calculated by each ACE potential fitted to each method and experimental results [51,52].

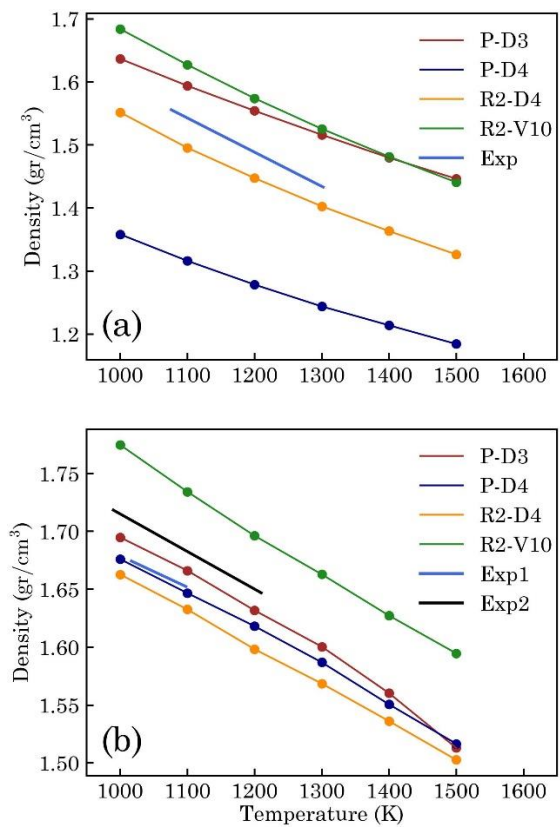

*Figure 8. Calculated densities of (a) NaCl, and (b) MgCl$_2$ using PBE-D3 (P-D3), PBE-D4 (P-D4), R2SCAN-D4 (R2-D4), R2SCANM-rVV10 (R2-V10) and experimental results (Exp)*

As seen in Figure 8, for NaCl the predictions of PBE-D3 and R2SCAN-rVV10 are rather close to each other and they both overestimate the density. The R2SCAN-D4 method underestimates the density of NaCl but it is closer to the experimental results compared to other methods. The PBE-D4 method severely underestimates the density of NaCl. For MgCl$_2$ we found two experimental data. The R2SCAN-rVV10 method overestimates both results, PBE-D3 falls between both experiments, PBE-D4 is very close to one of the experimental results and R2SCAN-D4 underestimates both experimental curves.

*Table 5. Comparison of properties calculated by PBE-D3 (P-D3), PBE-D4 (P-D4), R2SCAN-D4 (R2-D4), and R2SCAN-rVV10 (R2-V10) for NaCl and $MgCl_2$ at 1200 K and 1100 K respectively. The units are $gr/cm^3$ for density, J/(mol.K) for Specific heat, 1/K for thermal expansion coefficient (TEC) and $10^6$ $cm^2/s$ for diffusion coefficient (_diff). The experimental TECs are calculated using the density equations provided in the respective Refs. The uncertainties of the MD calculations are negligible (at most $3\times10^{-4}$ $gr/cm^3$ and on average $2\times10^{-4}$ $gr/cm^3$ for the densities, at most 0.29 J/mol.K and on average 0.16 J/mol.K for the specific heats, at most $0.03\times10^{-4}$ 1/K and on average $0.02\times10^{-4}$ 1/K for the thermal expansion coefficients, and at most 1% and on average 0.4% for the diffusivities)*

|  | P-D3 | P-D4 | R2-D4 | R2-V10 | Exp1 | Exp2 | Ref. |
|---|---|---|---|---|---|---|---|
| NaCl (1200 K) |  |  |  |  |  | - |  |
| Density | 1.55 | 1.28 | 1.45 | 1.57 | 1.49±0.5% | - | [51] |
| Specific heat | 64 | 60 | 64 | 70 | 66.9 | - | [53] |
| TEC | 2.5 | 2.8 | 3.2 | 3.2 | 3.7 | - | [51] |
| Na_diff | 119 | 159 | 127 | 114 | 105±20% | - | [54] |
| Cl_diff | 93 | 134 | 104 | 85 | 84±20% | - | [54] |
| $MgCl_2$ (1100 K) |  |  |  |  |  |  |  |
| Density | 1.67 | 1.65 | 1.63 | 1.73 | 1.65±1% | 1.68 | [51] , [52] |
| Specific heat | 94 | 93 | 94 | 96 | 92.0±0.1% |  | [55] |
| TEC | 1.9 | 1.8 | 2.1 | 2.2 | 1.6 | 1.9 | [51] , [52] |
| Mg_diff | 44 | 44 | 36 | 38 | - | - | - |
| Cl_diff | 48 | 48 | 41 | 40 | - | - | - |

In Table 5 we compare some thermophysical properties of NaCl and $MgCl_2$ calculated by the mentioned methods with the available experimental results. The value for each property calculated by MD simulations is the average value obtained from three independent simulations.

The uncertainties of the MD calculations are negligible (at most $3\times10^{-4}$ gr/cm$^3$ and on average $2\times10^{-4}$ gr/cm$^3$ for the densities, at most 0.29 J/mol.K and on average 0.16 J/mol.K for the specific heats, at most $0.03\times10^{-4}$ 1/K and on average $0.02\times10^{-4}$ 1/K for the thermal expansion coefficients, and at most 1% and on average 0.4% for the diffusivities), and we did not show them in the table. For NaCl, we chose to compare the data at 1200 K which falls in the middle of the experimentally available data. For MgCl$_2$ we chose 1100 K for the same reason. We could not find any experimental data for the diffusivity of MgCl$_2$. As can be seen in the table, the calculated properties by DFT-D4 for NaCl are far from other methods and seem to be unreliable. The other 3 methods, each perform better for a specific property. For MgCl$_2$, DFT-D4 shows a better performance followed by R2SCAN-D4 and PBE-D3 depending on which experimental results are more reliable. Overall, based on the results of Table 5, we cannot confidently select a single best-performing DFT method.

Figure 9 compares the calculated viscosities for the two salts using the four DFT methods and experimental results [51]. Each data point is the average of three independent calculations and the error bars show the standard deviation of the three calculations. In the NaCl simulations, once again the curve for PBE-D4 seems to be an unreliable result as it is significantly farther from the rest of the curves, so we set it aside. Among the other three methods, R2SCAN-D4 has the best performance followed by PBE-D3 and R2SCAN-rVV10. For MgCl$_2$ the difference between the methods is more apparent. The viscosities calculated by R2SCAN-rVV10 are very close to the experiments followed by R2SCAN-D4. PBE-D3 and PBE-D4 are much farther from the experimental curves. Based on our viscosity calculations we find that the R2SCAN-D4 method has consistently performed well. Adding this result to the results of Table 5, where R2SCAN-D4 performance was on par with the other methods, we select the R2SCAN-D4 to be the best overall approach in this study and suggest it might be a good starting point for future molten salt modeling. It should be noted that R2SCAN is a meta-GGA method and according to our timings, DFT calculations with R2SCAN are 5 times longer than PBE. The average running time of MD simulation with the developed ACE potentials is around $2\times10^{-4}$ (sec.core)/(atom.step) (tested on the recent AMD EPIC and INTEL XEON processors).

## 4. Discussion

In sections 3.1 and 3.2 we discussed two approaches to assessing the accuracy of fitting MLIPs for NaCl-MgCl$_2$ system as a representative binary molten salt: I) with respect to DFT energies and forces and predicted properties and II) with respect to experimentally measured properties. The following is the summary of our findings in this work:

1- It is possible to fit a compositionally transferable interatomic potential for binary molten salts using as low as 2500 training data by only including data from end members and two compositions in between them at every 33%. The data does not need to come from any AIMD simulations and the raw atomic configurations can initially be extracted from classical MD simulations or recently developed universal MLIPs [56] and be DFT calculated, followed by a few active learning cycles to increase the robustness of the potential.

2- In our study the fitted potential had energy and force errors, up to 4 meV/atom and 60 meV/Å (the highest errors in Figure 1). Such potential predicts material properties such as RDF, density, specific heat capacity, and thermal expansion coefficient with high accuracy compared to the underlying DFT, either within the errors of the DFT or with just about a 1% difference. We conclude that MLIPs could be surrogates for DFT methods and be used to assess the accuracy of the DFT methods with respect to the experimental results, up to the scales associated with these 1% and lower errors.

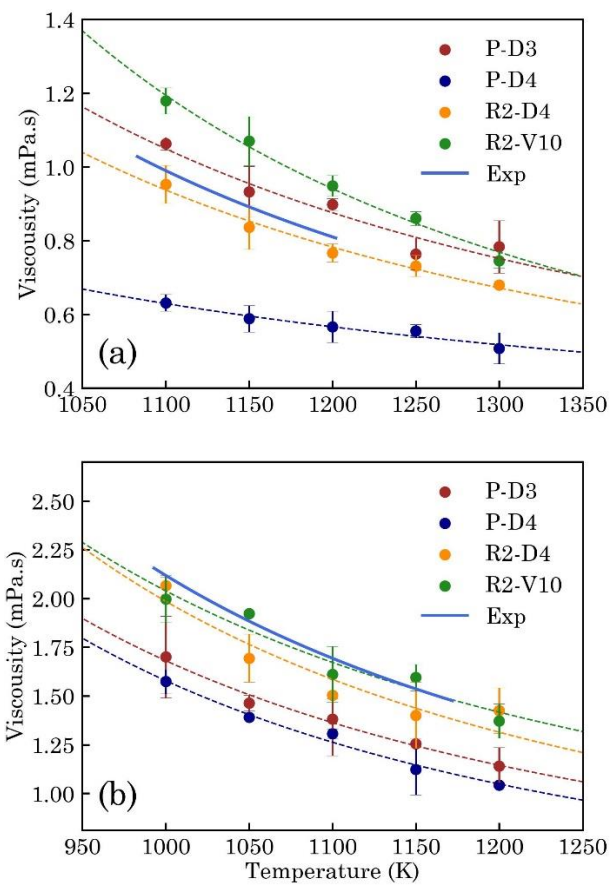

*Figure 9. Calculated viscosities of (a) NaCl, and (b) MgCl$_2$ using PBE-D3 (P-D3), PBE-D4 (P-D4), R2SCAN-D4 (R2-D4), R2SCANM-rVV10 (R2-V10) and experimental results (Exp)*

3- Based on our testing of four different DFT methods, namely PBE-D3, PBE-D4, R2SCAN-D4, and R2SCAN-rVV10, we conclude that R2SCAN-D4 has the best overall accuracy for the NaCl-MgCl2 system, although it is not a consistently better potential for all properties. Further salt studies with R2SCAN-D4 across different salt systems will be useful to assess its capabilities. While R2SCAN is a more computationally expensive method compared to PBE, the computational cost mostly affects simulations such as AIMD which are long and sequential. In these studies, based on DFT calculated values of distinct atomic configurations, R2SCAN was about 5 times slower than PBE.

## 5. Conclusions

In this work, we developed a compositionally transferable machine-learning interatomic potential for NaCl-MgCl$_2$ molten salt using the ACE potential and PBE-D3 method. We showed that by as low as 2500 training data and by only including training data from two end member salts and two compositions in between them at every 33%, one can develop such potential. The potential performed very well in reproducing DFT-calculated material properties of NaCl, MgCl$_2$, and eutectic NaCl-MgCl$_2$ ((NaCl)$_{0.58}$(MgCl$_2$)$_{0.42}$) suggesting that ACE potential could be used as a surrogate to DFT. We then fitted several other ACE potentials for unary NaCl and MgCl$_2$ using PBE-D4, R2SCAN-D4, and R2SCAN-rVV10 methods and assessed the accuracy of their property prediction compared to experimental data. Our results suggest that R2SCAN-D4 is overall a somewhat better method for reproducing the thermophysical properties of NaCl and MgCl$_2$ and we suggest that further study with this method would be useful.

**Data availability**

All the training and testing data, the fitted potentials, and the data to reproduce Figures 1 to 9 are provided in https://doi.org/10.6084/m9.figshare.26535952

**Acknowledgment**
We gratefully acknowledge support from the Department of Energy (DOE) Office of Nuclear Energy's (NE) Nuclear Energy University Programs (NEUP) under award # 21-24582. This work used Bridges-2 cluster at Pittsburgh Supercomputing Center (PSC) and Stampede3 cluster at Texas

Advanced Computing Center (TACC) through allocation MAT240071 from the Advanced Cyberinfrastructure Coordination Ecosystem: Services & Support (ACCESS) program [57], which is supported by National Science Foundation grants #2138259, #2138286, #2138307, #2137603, and #2138296. We also used the computational resources provided by the Center for High Throughput Computing (CHTC) at the University of Wisconsin–Madison.**References**

Innovation: NSF's Advanced Cyberinfrastructure Coordination Ecosystem: Services & Support, PEARC 2023 - Comput. Common Good Pract. Exp. Adv. Res. Comput. (2023) 173–176. https://doi.org/10.1145/3569951.3597559.

# Best Practices for Fitting Machine Learning Interatomic Potentials for Molten Salts: A Case Study Using NaCl-MgCl$_2$

*Siamak Attarian, Chen Shen, Dane Morgan, Izabela Szlufarska*

**Table of Contents:**

**Calculating the uncertainty of density due to energy errors:**

Basic definitions:

$$P_{V_0} = 0 \ (V_0 \text{ is the equilibrium volume}), \qquad \frac{dE}{dV}\bigg|_{V_0} = 0, \qquad B = -V\frac{dp}{dV} \ ; \ P = -\frac{dE}{dV}$$

$$B_0 = -V_0 \frac{dP}{dV}\bigg|_{V_0} = V_0 \frac{d^2E}{dV^2}\bigg|_{V_0} \implies \frac{d^2E}{dV^2}\bigg|_{V_0} = \frac{B_0}{V_0} \quad (B_0 \text{ is the bulk modulus at the equilibrium volume})$$

Taylor expansion of energy:

$$E(V) \cong E_0 + \left(\frac{\partial E}{\partial V}\bigg|_{V_0}\right)(V - V_0) + \frac{1}{2}\left(\frac{\partial^2 E}{\partial V^2}\bigg|_{V_0}\right)(V - V_0)^2$$

The first derivative is zero:

$$E(V) \cong E_0 + \frac{1}{2}\left(\frac{\partial^2 E}{\partial V^2}\bigg|_{V_0}\right)(V - V_0)^2$$

Bulk modulus of NaCl is around 3 GPa at 1200 K [1]:

$$B_0 = 3 GPa = 0.19 \ ev/Å^3$$

Replacing the second derivative:

$$E(V) \cong E_0 + \frac{1}{2}\left(\frac{B_0}{V_0}\right)(V - V_0)^2 = E_0 + \frac{1}{2}\left(\frac{0.19}{V_0}\right)(V - V_0)^2$$

$$\delta E \cong \frac{1}{2}\left(\frac{0.19}{V_0}\right)(\delta V)^2 = \left(\frac{0.095}{V_0}\right)(\delta V)^2 \implies \delta V = \sqrt{\frac{\delta E \times V_0}{0.095}} = \sqrt{10.52 \times \delta E \times V_0} = 3.24\sqrt{\delta E \times V_0};$$

For NaCl at 1200 K, from our AIMD calculations: $V_0$ = 30.83 Å$^3$/atom and the average mass of an atom of (Na or Cl) is 4.85×10$^{-23}$ gr (Density of NaCl: 1.576 gr/cm$^3$ from AIMD )

For $\delta E = 0.002 \ eV/atom$: $\delta V = 3.24\sqrt{0.002 \times 30.83} = 0.8 \ Å^3 = 0.8 \times 10^{-24} \ cm^3/atom$

$$\delta\rho = \frac{\delta V}{M} = \frac{0.8 \times 10^{-24}}{4.85 \times 10^{-23}} = 0.017 \frac{gr}{cm^3}$$

So, the percent error in density corresponding to a 2 meV/atom error in energy is

$$\frac{0.017}{1.576} \times 100 \cong 1.1\% \; error$$

Table S.1. List of compositions and their %NaCl-%MgCl$_2$ ratio

| # | NaCl-MgCl$_2$ | # | NaCl-MgCl$_2$ |
|---|---|---|---|
| #1 | 100%-0% | #18 | 59%-41% |
| #2 | 98%-2% | #19 | 55%-45% |
| #3 | 96%-4% | #20 | 52%-48% |
| #4 | 94%-6% | #21 | 49%-51% |
| #5 | 91%-9% | #22 | 46%-54% |
| #6 | 89%-11% | #23 | 42%-58% |
| #7 | 87%-13% | #24 | 39%-61% |
| #8 | 85%-15% | #25 | 35%-65% |
| #9 | 82%-18% | #26 | 32%-68% |
| #10 | 80%-20% | #27 | 28%-72% |
| #11 | 77%-23% | #28 | 25%-75% |
| #12 | 75%-25% | #29 | 20%-80% |
| #13 | 72%-28% | #30 | 17%-83% |
| #14 | 70%-30% | #31 | 12%-88% |
| #15 | 67%-33% | #32 | 9%-91% |
| #16 | 64%-36% | #33 | 3%-97% |
| #17 | 61%-39% | #34 | 0%-100% |

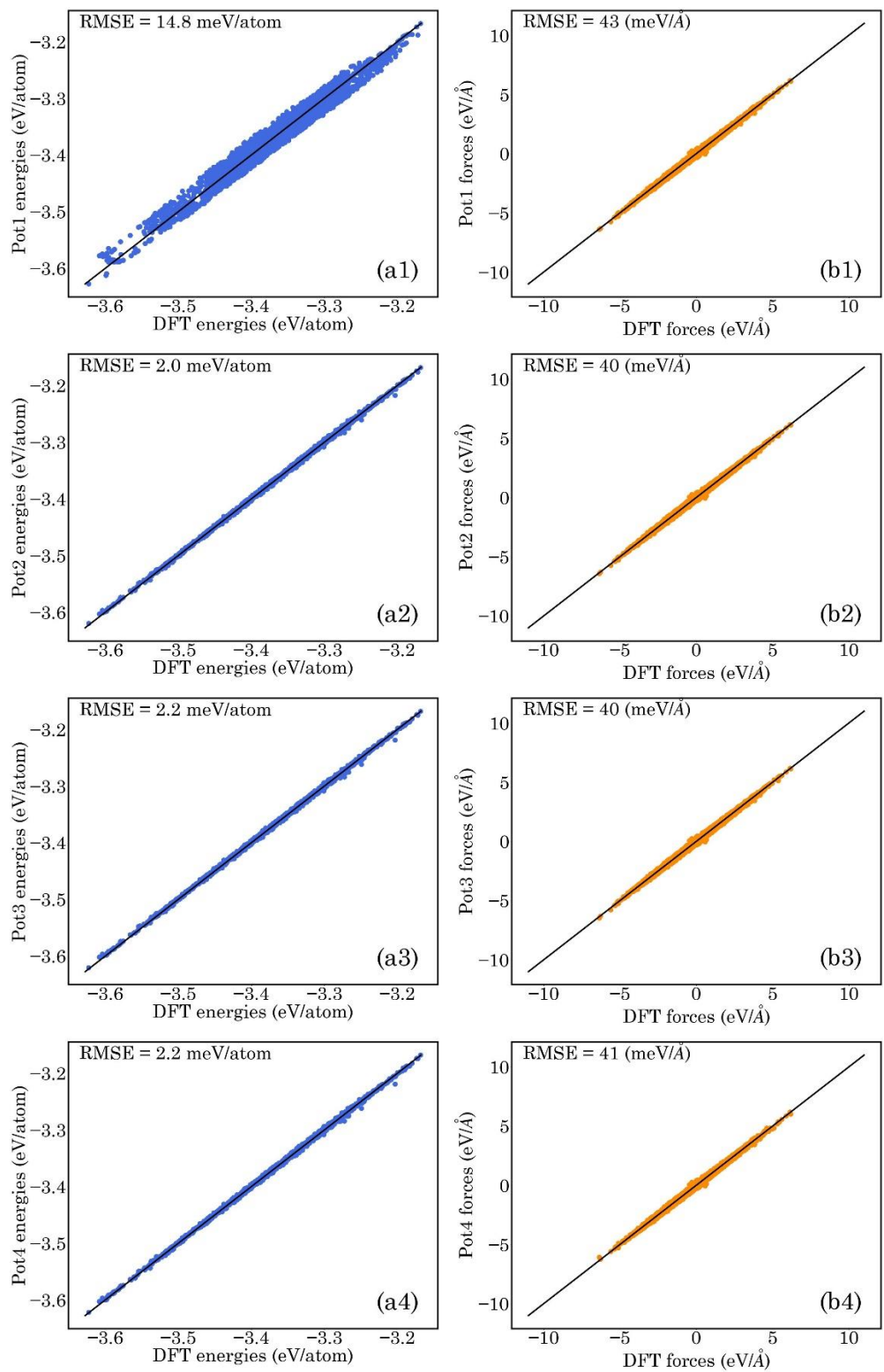

*Figure S1. Parity plots of energies(a1 to a4) and forces (b1 to b4) of testing set predicted by Pot_1 to Pot_4 and DFT.*

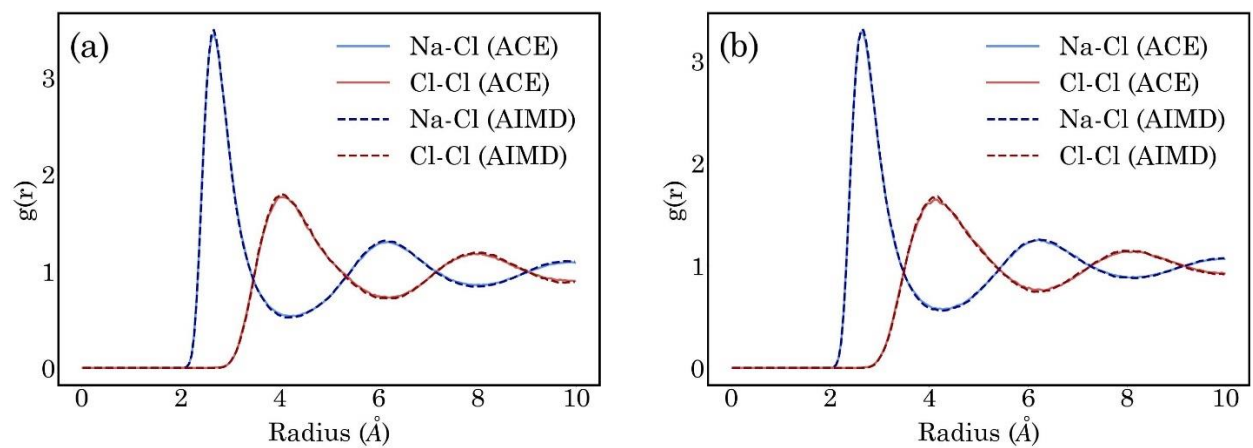

*Figure S2. Radial distribution functions (RDF or g(r)) of Na-Cl, and Cl-Cl pairs in pure NaCl salt at temperatures (a) 1200 K, and (b) 1500 K calculated from ACE and AIMD.*

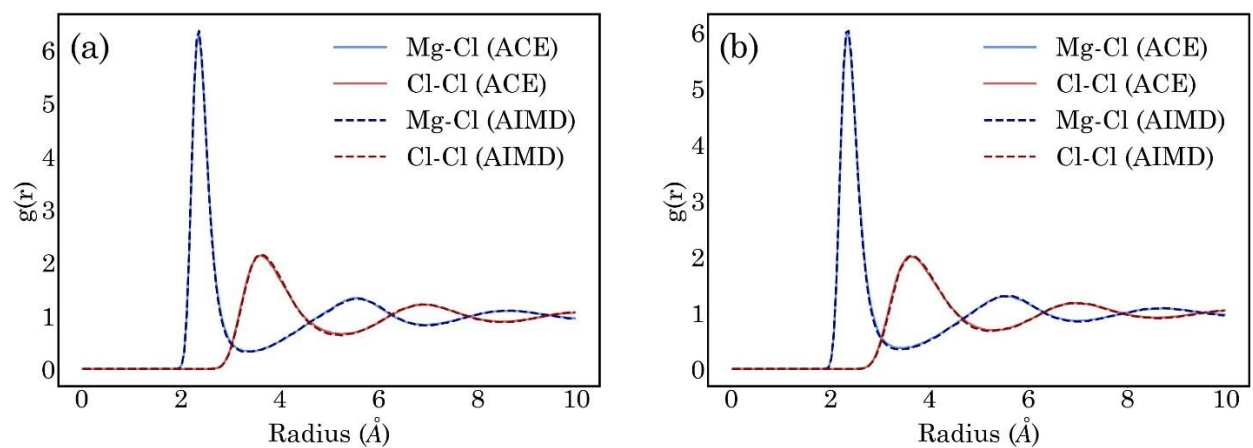

*Figure S3. Radial distribution functions (RDF or g(r)) of Mg-Cl, and Cl-Cl pairs in pure MgCl$_2$ salt at temperatures (a) 1200 K, and (b) 1500 K calculated from ACE and AIMD.*

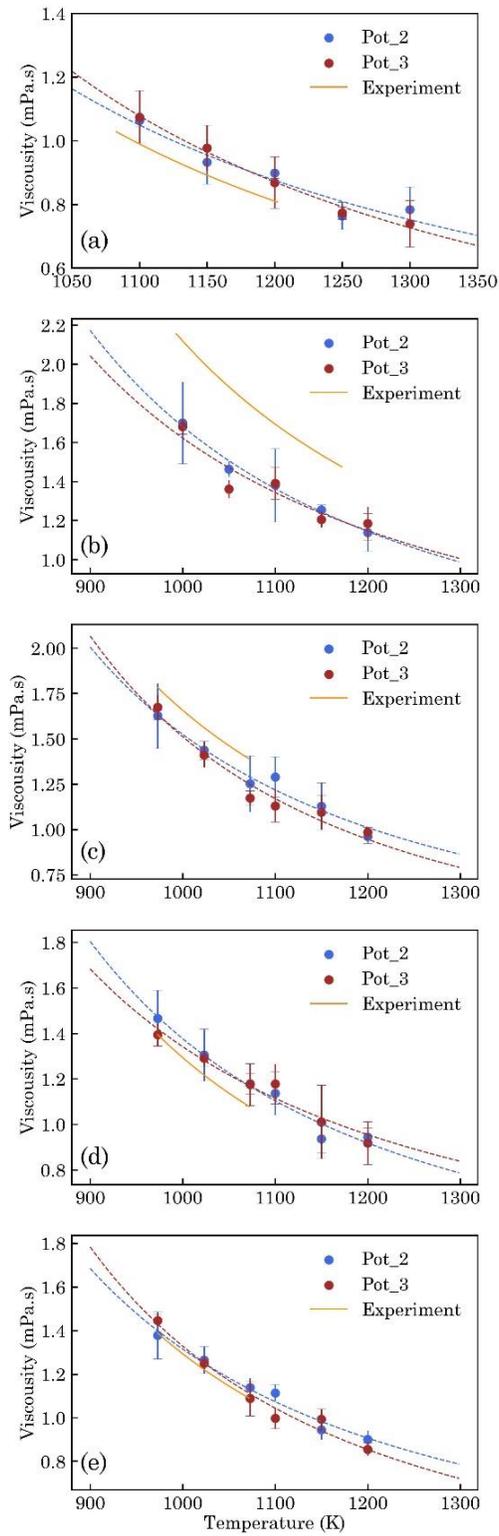

*Figure S4. Viscosities of (a) NaCl, (b) MgCl$_2$, (c) compo4, (d) compo5 and, (e) compo6 calculated by Pot_2 and Pot_3 compared to experiments.*

In Figure S4 we see that the results of Pot_2 and Pot_3 are very close mostly within each other's error bars. This figure shows that the deviation of the predicted viscosities from experimental results by either of these potentials is rather similar and both potentials could have represented the PBE-D3 method.

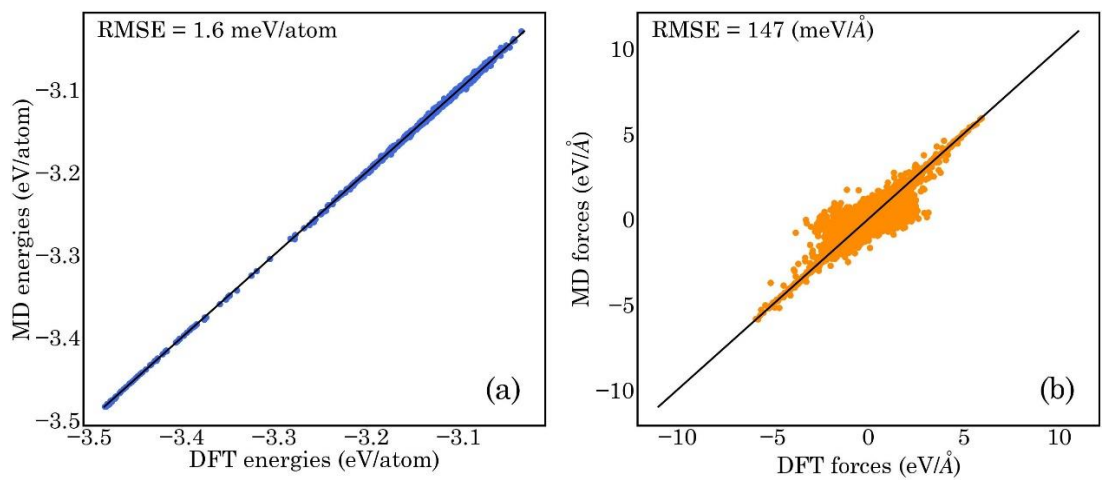

*Figure S5. Parity plots of energies(a) and forces (b) of the training set predicted by the potential fitted to the PBE-D4 data (MD_forces) and true PBE-D4 (DFT_forces).*

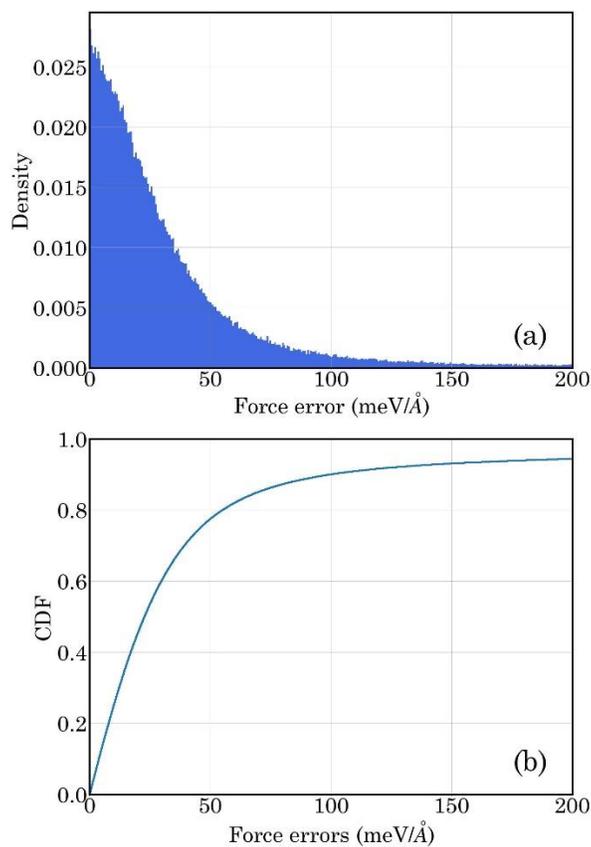

*Figure S6. Histogram and cumulative density function of the force errors of the training set predicted by the potential fitted to the PBE-D4 data.*

In Figure S5 the energy errors predicted by the potential that is fitted to PBE-D4 data are low while the force errors are much higher than what we get doing the same procedure of fitting for PBE-D3. Due to the huge number of atomic forces presented in Figure S5(b), it is hard to show the fact that the majority of points fall on top of or very close to the ideal line and the points that are farther look more prominent. In Figure S6 we show the distribution of force errors. As can be seen, 90% of the force errors are below 100 meV/Å. As discussed in the paper, increasing the complexity of the ACE potential did not help in bringing the scattered points in Figure S5(b) closer to the ideal line while many points that were rather close to the ideal line, would get even closer by increasing the complexity of the potentials. In Figure S7 we compare the calculated forces by the PBE-D4, R2SCAN-D4, and R2SCAN-rVV10 methods to the PBE-D3 method. Since each method has its assumptions we expect some degree of difference between the calculated forces by each method, but overall, we expect a general agreement between the calculations. We quantified the difference between the calculated forces by the parameter root mean squared difference (RMSD) in the plots which is basically the standard deviation between the force calculations. In Figure S7(a) we see a noticeable scatter meaning that some calculated forces by the PBE-D4 method are much different than PBE-D3. The severity of the scatter is much less between PBE-D3 and R2SCAN-D4 in Figure S7(b), and in Figure S7(c) the forces calculated by PBE-D3 and R2SCAN-rVV10 agree well with minor differences. The same plots are shown for $MgCl_2$ in Figure S8 and as can be seen the force calculations agree well between the methods. We concluded that these scattered points may have been miscalculated and there may be some issues in the way that PBE-D4 dispersion correction is being calculated for the Na ion.

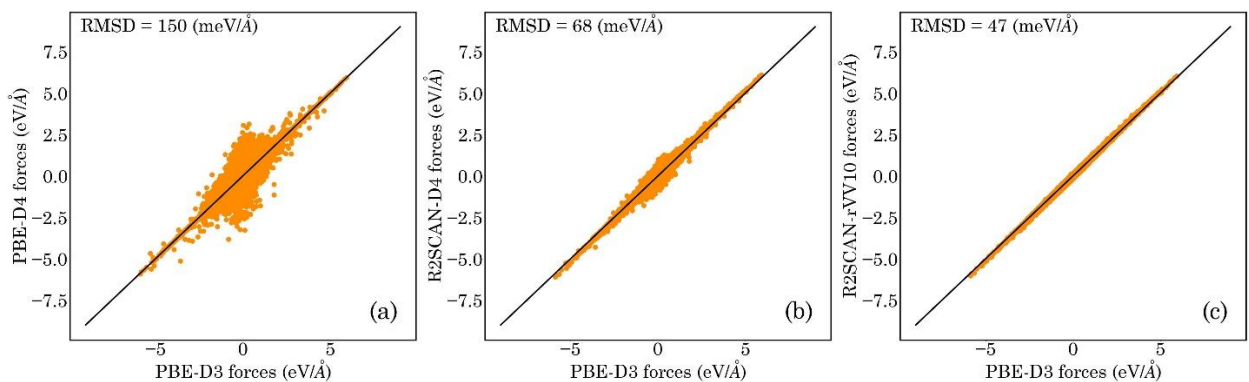

*Figure S7. Comparing the calculated forces by PBE-D4 (a), R2SCAN-D4 (b), and R2SCAN-rVV10 (c) methods with the forces calculated by the PBE-D3 method for the NaCl system.*

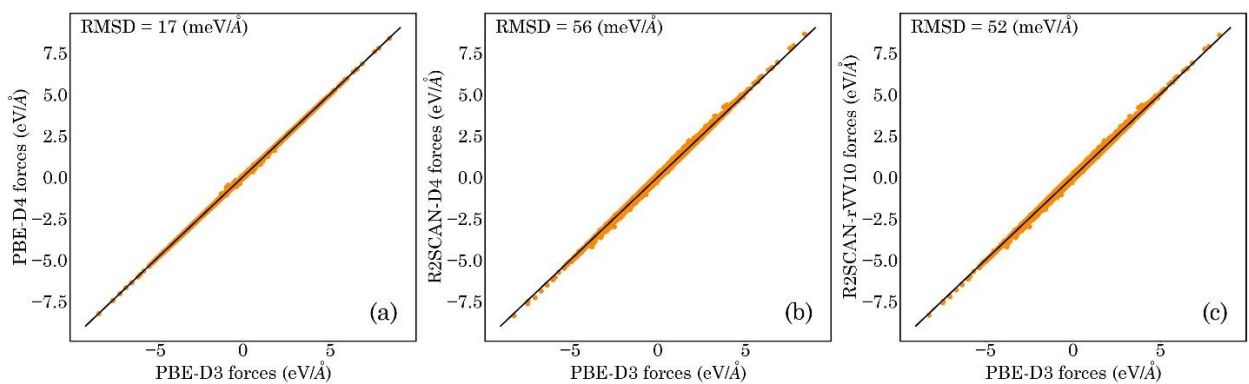

*Figure S8. Comparing the calculated forces by PBE-D4 (a), R2SCAN-D4 (b), and R2SCAN-rVV10 (c) methods with the forces calculated by the PBE-D3 method for the MgCl2 system.*